\documentclass[twocolumn,epjc3,a4paper]{svjour3}

\usepackage[numbers,sort&compress]{natbib}

\usepackage[utf8]{inputenc}
\usepackage[english]{babel}
\usepackage{amssymb,amsmath,amstext,amsbsy,amsopn}
\usepackage{bbm}
\usepackage{graphicx}
\usepackage{tabularx}
\usepackage{lmodern}

\usepackage{microtype}

\journalname{Eur. Phys. J. A}
\renewcommand{\email}[1]{e-mail: \href{mailto:#1}{#1}}

\usepackage[
    colorlinks=true,
    linkcolor=blue,
    citecolor=blue,
    urlcolor=blue
]{hyperref}
\usepackage{orcidlink} 

\newcommand{\NNN}{\text{3N}}

\begin{document}

\allowdisplaybreaks

\title{Neutron-rich nuclei and neutron skins from chiral low-resolution interactions}

\author{P.~Arthuis\thanksref{1,2,3,aff:pa}\orcidlink{0000-0002-7073-9340} \and K.~Hebeler\thanksref{1,3,4,em:kh}\orcidlink{0000-0003-0640-1801} \and A.~Schwenk\thanksref{1,3,4,em:as}\orcidlink{0000-0001-8027-4076}}

\institute{%
Technische Universit\"at Darmstadt, Department of Physics, 64289 Darmstadt, Germany \label{1} 
\and Helmholtz Forschungsakademie Hessen für FAIR (HFHF), GSI Helmholtzzentrum für Schwerionenforschung GmbH, 64291 Darmstadt, Germany \label{2}
\and ExtreMe Matter Institute EMMI, GSI Helmholtzzentrum f\"ur Schwerionenforschung GmbH, 64291 Darmstadt, Germany \label{3}
\and Max-Planck-Institut f\"ur Kernphysik, Saupfercheckweg 1, 69117 Heidelberg, Germany \label{4}}

\thankstext{aff:pa}{Current affiliation: Universit\'e Paris-Saclay, CNRS/IN2P3, IJCLab, 91405 Orsay, France.\\
\email{pierre.arthuis@ijclab.in2p3.fr}}
\thankstext{em:kh}{\email{kai.hebeler@physik.tu-darmstadt.de}}
\thankstext{em:as}{\email{schwenk@physik.tu-darmstadt.de}}

\date{}


\maketitle

\begin{abstract}
Neutron-rich nuclei provide important insights to nuclear forces and to the nuclear equation of state. Advances in \emph{ab initio} methods combined with new opportunities with rare isotope beams enable unique explorations of their properties based on nuclear forces applicable over the entire nuclear chart. In this paper, we develop novel chiral low-resolution interactions that accurately describe bulk properties from $^{16}$O to $^{208}$Pb. With these, we investigate density distributions and neutron skins of neutron-rich nuclei. Our results show that neutron skins are narrowly predicted over all nuclei with interesting sensitivities for the most extreme, experimentally unexplored cases.
\end{abstract}

\section{Introduction}

Neutron-rich nuclei constitute natural laboratories at the intersection of nuclear physics and astrophysics. Away from the valley of stability, shell structure evolves, and neutron-rich nuclei become sensitive to important parts of nuclear forces~\cite{Hebe15ARNPS,Otsuka2020,Nowacki2021}. The larger isospin asymmetry also offers natural connections to the nuclear equation of state and infinite matter in neutron stars. At the most neutron-rich extremes, nuclei are key for the r-process and nucleosynthesis~\cite{Schatz2022}. As a result, neutron-rich nuclei constitute the main science driver for rare isotope beam facilities such as FAIR, FRIB, and RIBF. Simultaneously, advances in \emph{ab initio} methods~\cite{Herg20review,Hebe203NF} extended their reach up to $^{208}$Pb~\cite{Miyagi2021,Hu22Pb,Hebeler2022jacno}. This offers tremendous perspectives of mutual feedback between theoretical and experimental efforts in the years to come.

Neutron skins have been a key focus for neutron-rich nuclei.
In particular, the recent extraction of a large $^{208}$Pb neutron skin from parity-violating electron scattering~\cite{Adhikari2021prex2}, albeit with large uncertainties, has sparked intense discussions within the nuclear physics and astrophysics communities (see, e.g., Refs.~\cite{Essick2021,Hu22Pb,Reinhard2022,Mammei2023}).
Through their correlations with the nuclear equation of state, neutron skins constitute an important connection between nuclei and dense matter in neutron stars.
Because the neutron skin increases with isospin asymmetry, the upcoming experimental access to many unexplored neutron-rich nuclei paves the way for discoveries critical to our understanding of both finite and infinite nuclear systems.

The description of atomic nuclei from first principles is based on the development of nuclear interactions, many-body methods, and advances in scientific computing.
The use of interactions based on chiral effective field theory~\cite{Epel09RMP,Entem2020,Epelbaum2020,Hammer2020EFT} and polynomially scaling many-body methods~\cite{Hage14RPP,Soma14GGF2N3N,Herg16PR,Tichai2020review} has allowed for a rapid extension of the reach of \emph{ab initio} methods both in terms of mass number $A$ and of observables of interest.
For applications across the nuclear chart, low-resolution interactions, either fit with a low-momentum cutoff $\Lambda$ or obtained from a similarity renormalization group (SRG) evolution~\cite{Bogn07SRG,Furn13RPP} have proven especially successful~\cite{Hebe11fits,Ekst15sat,Huth19chiralfam,Soma2020,Jian20N2LOGO}.
In particular, they offer good convergence properties with respect to both model space size and expansion order of the many-body method.
This makes them key for exploring nuclei at the frontier of the current \emph{ab initio} reach.
In particular, the 1.8/2.0~(EM) two- and three-nucleon (NN+3N) interaction~\cite{Hebe11fits} has been widely used for studies of ground and excited states up to heavy nuclei~\cite{Morr17Tin,Stroberg2021,Miyagi2021,Hebeler2022jacno,Tichai2023}, but suffers from underprediction of radii~\cite{Simo17SatFinNuc}.
While Delta-full interactions have been successful in this respect~\cite{Jian20N2LOGO,Hu22Pb} by fitting to low-energy NN phase shifts and many-body data at low cutoffs, Delta-less interactions have had difficulties at reproducing simultaneously energies and radii~\cite{Simo17SatFinNuc,Soma20SCGF}, or have lacked the capacity to reach past the tin isotopes~\cite{Ekst15sat,Huth19chiralfam}. In addition, attempts based on consistent SRG evolution of the two- and three-body sectors have exhibited difficulties at providing accurate results for medium-mass and heavy nuclei and the need for refits to many-body data~\cite{Hopp19medmass,Huth19chiralfam}.
As such, it appears critical to develop low-resolution interactions with good convergence properties and yielding a good reproduction of bulk properties of nuclei.

\section{Novel low-resolution interactions}

To this end, we explore low-resolution NN+3N interactions following the strategy of Ref.~\cite{Hebe11fits}. Specifically, we start from a NN interaction that is evolved to lower resolution scales ($\lambda = 1.8\,\mathrm{fm}^{-1}$) via the SRG, which leaves all two-body observables unchanged by construction and preserves the accurate reproduction of NN phase shifts~\cite{Bogn10PPNP}. In a second step, the shorter-range 3N interactions at next-to-next-to-leading order (N$^2$LO) are fitted at low resolution with a cutoff $\Lambda_\NNN = 2.0\,\mathrm{fm}^{-1}$ in combination with the SRG-evolved NN interaction to few- and many-body observables.\footnote{For additional details see \ref{app:manybody}.} The long-range LECs $c_1$, $c_3$, and $c_4$ ($c_i$) of the 3N interactions are kept consistent with the corresponding NN potentials. This strategy assumes that the long-range parts of nuclear forces are unchanged by the SRG evolution, and that the chiral interactions remain an adequate operator basis at low resolution. In addition, we constrain the shorter-range 3N LECs $c_D$ and $c_E$ to reproduce the $^{3}$H ground-state energy.\footnote{See \ref{app:fewbody} for reproduction of few-body systems.} This choice leads to excellent many-body convergence for medium-mass to heavy nuclei~\cite{Simo17SatFinNuc,Morr17Tin,Miyagi2021,Hebeler2022jacno}. Moreover, this strategy has led to the 1.8/2.0 (EM) NN+3N interaction~\cite{Hebe11fits}, which exhibits remarkable agreement with experimental binding energies over a wide range of the nuclear chart (see, e.g., Refs.~\cite{Simo17SatFinNuc,Stroberg2021}).

Despite the successes of interactions fitted at low resolution, the Hamiltonians derived in Ref.~\cite{Hebe11fits} also exhibit shortcomings, in particular in form of too small charge radii of medium-mass and heavy nuclei compared to experiment. Hence, in this work we generalize this approach by extending the set of considered NN interactions and by including many-body observables in the fits. Specifically, we consider in addition to the N$^3$LO EM 500 NN potential~\cite{Ente03EMN3LO} two more recent families of NN interactions: First, we explore the NNLOsim family of N$^2$LO potentials~\cite{Carl15sim}, where we focus on the interactions optimized to scattering energies up to $T^\mathrm{max}_\mathrm{Lab} = 290$\,MeV. Second, we use the EMN potentials of Ref.~\cite{Ente17EMn4lo} at N$^2$LO, consistent with the order of the employed 3N interactions. For the EMN potentials the $c_i$ LECs are taken from the Roy-Steiner analysis~\cite{Hofe15piNchiral} of pion-nucleon scattering data, while for the NNLOsim potentials they are optimized to NN and 3N data. For both families of interactions, we consider the three cutoff values $\Lambda_{\text{NN}} = 450, 500$, and 550\,MeV.

\begin{figure}[t!]
  \includegraphics[width=\linewidth]{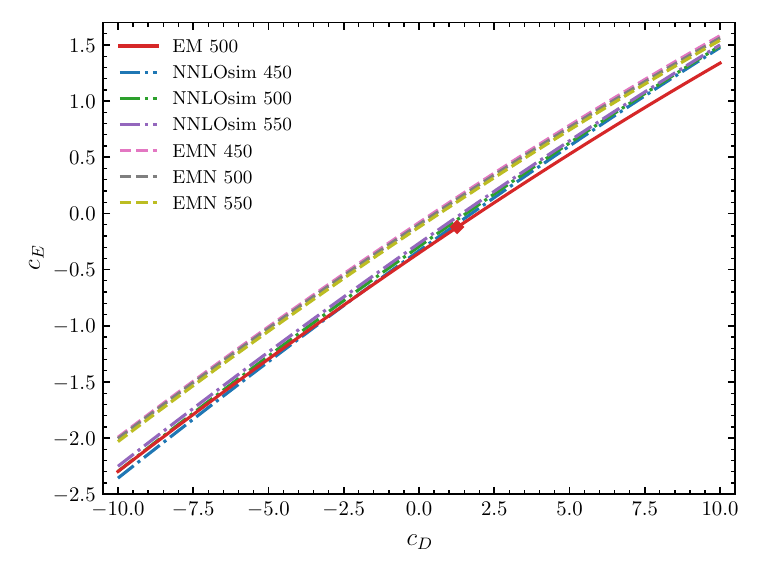}%
  \caption{\label{fig:triton_fit} Three-nucleon LECs $c_D$ and $c_E$ that reproduce the $^3$H binding energy for the employed low-resolution NN+3N interactions at NN SRG resolution scale $\lambda = 1.8 \, \text{fm}^{-1}$. The coupling values of the 1.8/2.0~(EM) interaction~\cite{Hebe11fits} are indicated by a red diamond for reference.}%
  \label{fig:3Hfits}%
\end{figure}

As in Ref.~\cite{Hebe11fits}, we first constrain the two LECs $c_D$ and $c_E$ to the experimental binding energy of $^3$H by solving the bound-state Faddeev equations. Figure~\ref{fig:3Hfits} shows the results for the low-resolution NN+3N interactions discussed above (with the NN potential SRG-evolved to the resolution scale $\lambda = 1.8 \, \text{fm}^{-1}$), as well as for the 1.8/2.0 (EM) interaction~\cite{Hebe11fits}. The relations between the two LECs are almost linear, indicating the perturbativeness of these low-resolution interactions. Furthermore, within a given family of NN interactions, the obtained trajectories are very close to each other. This is reminiscent of the NN universality at low resolution~\cite{Bogn10PPNP}, so that the impact of the cutoff of the unevolved NN potential is small. In particular, we find almost identical results for the three EMN interactions, which all share the same $c_i$ LECs. This demonstrates a high degree of universality of SRG-evolved NN interactions at this low resolution scale.

All many-body results presented in this work are calculated using the \emph{ab initio} in-medium similarity renormalization group (IMSRG)~\cite{Herg16PR} in the IMSRG(2) approximation, i.e., at the normal-ordered two-body level based on a Hartree-Fock reference state. The nuclear Hamiltonian is represented in a spherical harmonic-oscillator basis of size $e_\mathrm{max} = \mathrm{max}(2n + l) = 14$, where $n$ is the radial quantum number and $l$ the orbital angular momentum. Matrix elements of the 3N forces are represented in a basis defined by $e_1 + e_2 + e_3 \leq E^{(3)}_\mathrm{max}$ with $E^{(3)}_\mathrm{max} = 24$~\cite{Miyagi2021}. The harmonic-oscillator frequency $\hbar\omega$ is selected to yield the best converged result depending on the system and observable, and ranges from 10 to 16\,MeV. 

\begin{figure}[t!]
  \includegraphics[width=\linewidth]{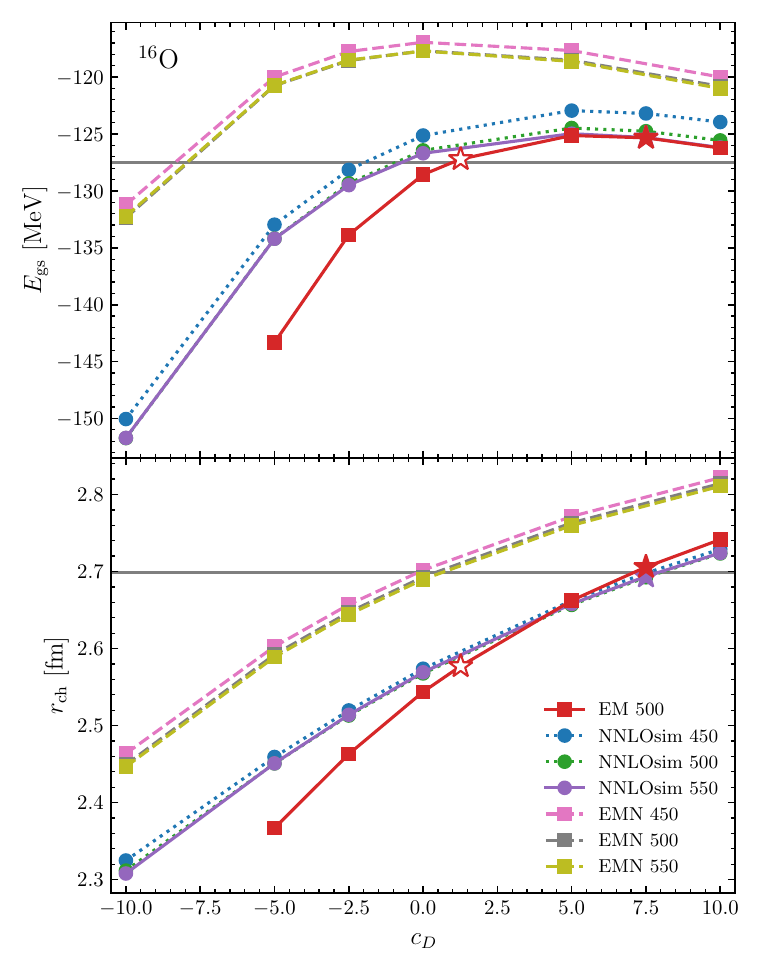}%
  \caption{\label{fig:O16_Egs} Ground-state energy (top panel) and charge radius (bottom panel) of $^{16}$O as a function of $c_D$ for the low-resolution NN+3N interactions at NN SRG resolution scale $\lambda = 1.8 \, \text{fm}^{-1}$. Horizontal grey lines correspond to the experimental values~\cite{Wang2021AME,Ange13rch}. The solid stars indicate the results corresponding to the two novel interactions 1.8/2.0 (EM7.5) and 1.8/2.0 (sim7.5), while the open star corresponds to the 1.8/2.0 (EM) interaction~\cite{Hebe11fits}.}
\end{figure}

Figure~\ref{fig:O16_Egs} shows the ground-state energy and charge radius of $^{16}$O as a function of $c_D$ for the different low-resolution NN+3N interactions. In all cases, we find a rather strong sensitivity of the ground-state energy (upper panel) for negative $c_D$ values, whereas for positive $c_D$ the energy is rather insensitive to changes of the LEC. While the NNLOsim family and EM 500 potential show a very good agreement with experiment at positive $c_D$, the EMN family exhibits larger deviations. For the charge radius (lower panel), we find an almost linear dependence and increasing radii for larger $c_D$ values. In particular, around $c_D$ = 7.5 the SRG-evolved NNLOsim and EM 500 interactions simultaneously reproduce both observables very well. On the other hand, the EMN interactions reproduce the experimental charge radius of $^{16}$O for a value of $c_D \sim$  0. However, here it is not possible to reproduce both observables simultaneously. Therefore, we selected two novel low-resolution interactions based on the EM 500 and NNLOsim~550 with a 3N LEC $c_D = 7.5$\footnote{While this value of $c_D$ appears particularly large, it is consistent with previous fits of SRG-evolved low-cutoff interactions~\cite{Huth19chiralfam}. However, due to the differences between the 1.8/2.0 approach and NN+3N SRG-evolution, we obtain different predictions for medium-mass nuclei based on the EMN NN family of interactions, as shown in \ref{app:manybody}. We were also not able to obtain fits using our strategy for the unevolved NN+3N interactions with the EMN interactions. This points to the fact that consistently evolved NN+3N interactions~\cite{Huth19chiralfam} are not unitarily equivalent to the bare ones, likely for lack of induced 4N forces.}, that we named 1.8/2.0~(EM7.5) and 1.8/2.0~(sim7.5), respectively. As can be seen in Fig.~\ref{fig:O16_Egs}, both these two interactions, indicated by a solid star, predict essentially the same binding energies as the 1.8/2.0~(EM) interaction (marked by an open star), but resolve the problem of too small charge radii.\footnote{ We compute the charge radius using the updated recommendation for the proton charge radius corresponding to $R_p^2 = 0.707\,\text{fm}^2$~\cite{Zyla2020}. This translates into a reduction of the charge radius of 0.012\,fm for $^{16}$O and 0.02\,fm for $^4$He, so very small compared to the total change.} This observation also holds for $^{40}$Ca, see \ref{app:manybody}.

\begin{figure}[t!]
  \includegraphics[width=\linewidth]{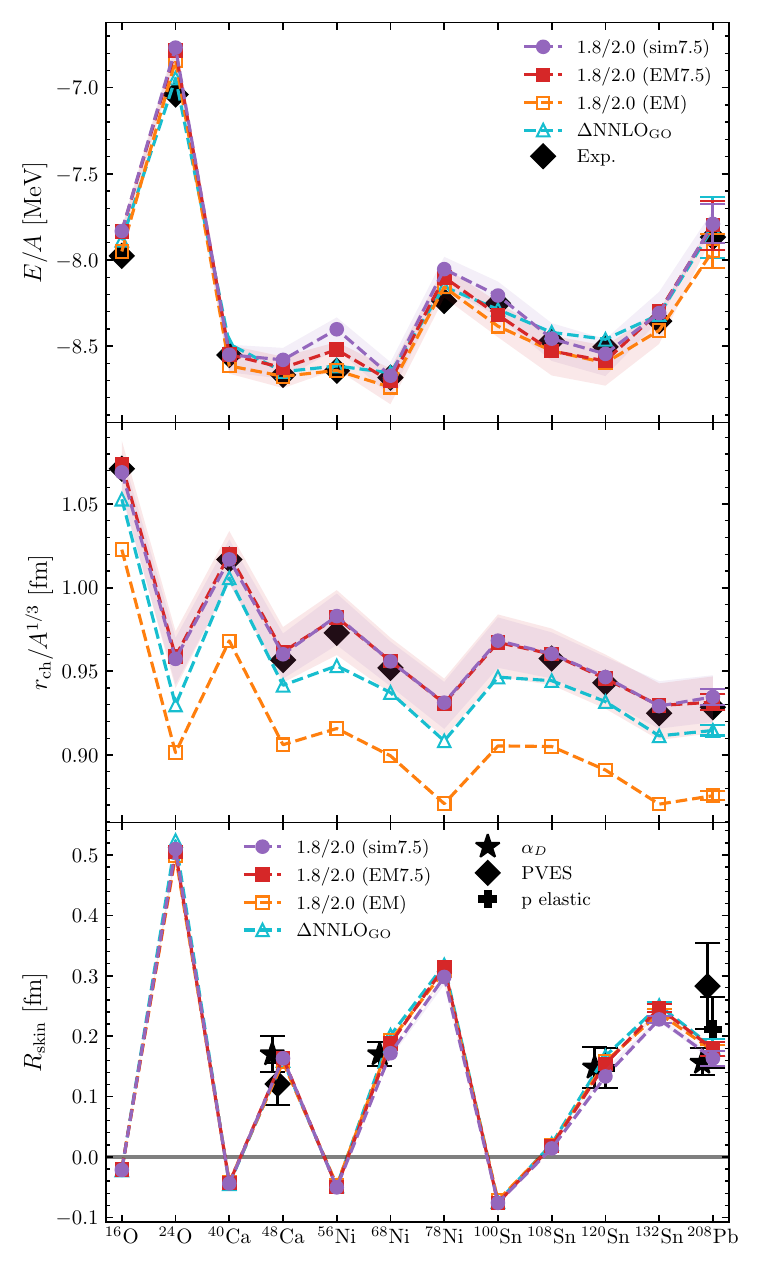}%
  \caption{\label{fig:Main_selected} Ground-state energies per nucleon, charge radii, and neutron skins for selected doubly closed-shell nuclei from $^{16}$O to $^{208}$Pb compared with experiment for the two novel interactions 1.8/2.0 (EM7.5) and 1.8/2.0 (sim7.5) as well as the 1.8/2.0 (EM) and $\Delta$NNLO$_\text{GO}$ interactions. Charge radii are scaled by $A^{1/3}$ for legibility. Experimental data for neutron skins were obtained through dipole polarizability ($\alpha_D$), parity-violating electron scattering (PVES), and elastic proton scattering (p elastic). See text for experimental references and for a discussion of the theoretical uncertainties given.}
\end{figure}

\section{Ground-state energies, radii, and neutron skins}

Figure~\ref{fig:Main_selected} shows the ground-state energy and charge radius for selected doubly closed-shell nuclei ranging from $^{16}$O to $^{208}$Pb based on our novel low-resolution interactions as well as the established 1.8/2.0~(EM)~\cite{Hebe11fits} and $\Delta$NNLO$_\mathrm{GO}$ ($\Lambda_\NNN = 394$\,MeV) interactions~\cite{Jian20N2LOGO}. For both 1.8/2.0~(EM7.5) and 1.8/2.0~(sim7.5), the bands are obtained by varying $c_D$ from 5 to 10 and represent an uncertainty estimate associated to our choice of $c_D$ value. The top panel displays ground-state energies per nucleon compared to experimental values from Ref.~\cite{Wang2021AME}. All interactions show good agreement over the entire range, with deviations from experiment below 2\%\footnote{The only exception is $^{56}$Ni with the 1.8/2.0~(sim7.5) interaction, where the deviation is associated to an anomalously low third-order many-body perturbation theory correction.}, the typical range of deviation between results from IMSRG(2) and exact methods~\cite{Herg17PS}.
In the case of $^{208}$Pb, computed energies were extrapolated assuming a geometric convergence with respect to $e_\mathrm{max}$\footnote{We obtain the extrapolated ground-state energy as follows, $E_\mathrm{gs} = E_\mathrm{gs}({e_\mathrm{max}=12}) + \delta E_\mathrm{gs}^{14}/[1 - (\delta E_\mathrm{gs}^{14} - \delta E_\mathrm{gs}^{12})]$ with $\delta E_\mathrm{gs}^{n} = E_\mathrm{gs}({e_\mathrm{max}=n}) - E_\mathrm{gs}({e_\mathrm{max}=n-2})$.}, and error bars correspond to change from $e_\mathrm{max} = 12$ to 14 to give an estimate of the converged value. The agreement with experiment is remarkably good, and thus establishes the new low-resolution interactions as very promising for \emph{ab initio} studies across the nuclear chart. The central panel displays the charge radius scaled by $A^{1/3}$. Here the new interactions display a strikingly good agreement with experiment~\cite{Ange13rch,MalbrunotEttenauer2022,Sommer2022}, and in particular a significant improvement with respect to the 1.8/2.0~(EM) and $\Delta$NNLO$_\mathrm{GO}$ interactions. Finally, the bottom panel shows the neutron skin, defined as the difference between point-neutron and point-proton radii, compared to experimental extractions from the dipole polarizability~\cite{Birk17dipole,Ross13dipole,Hash15dipole,Tami11dipole}, parity-violating electron scattering~\cite{Adhikari2022crex,Adhikari2021prex2}, and elastic proton scattering~\cite{Terashima2008,Zeni10neutrondistPb}. We find that the neutron skins are predicted remarkably similar for all interactions explored, even if the charge radii are underestimated from 1.8/2.0~(EM) and $\Delta$NNLO$_\mathrm{GO}$. This is likely because chiral low-resolution interactions tightly constrain the properties of neutron matter, which is key for the neutron skin as a difference of neutron and proton radii, while the individual radii do depend on the Hamiltonian. Our results also agree well with experiment, with the exception of the PREX result~\cite{Adhikari2021prex2} for $^{208}$Pb, where the agreement is only within the $2\sigma$ range. These results demonstrate the capability of our new low-resolution interactions for neutron skins, and thus can also be used in experimental analyses to provide constraints on the symmetry energy slope $L$~\cite{Auma17peelsym}.

\begin{figure}[t]
  \includegraphics[width=\linewidth]{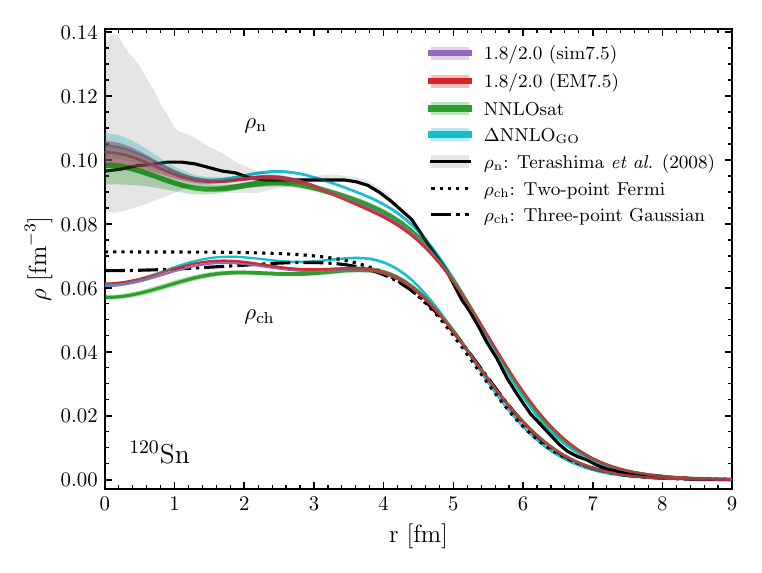}%
  \caption{\label{fig:Sn120_rho} Charge and point-neutron density distributions for $^{120}$Sn for the two novel interactions 1.8/2.0 (EM7.5) and 1.8/2.0 (sim7.5) as well as the NNLOsat and $\Delta$NNLO$_\text{GO}$ interactions, compared to experimental constraints~\cite{DeVries1987ADNDT,Terashima2008}. The darker and lighter theoretical uncertainty bands estimate the convergence by varying $\hbar\omega$ and $e_\mathrm{max}$, respectively.}
\end{figure}

\section{Density distributions}

Density distributions offer insights into nuclear structure by combining a bulk description of the nucleus with information on the internal shell. With ample information for stable nuclei, electron scattering experiments on rare isotopes are also becoming possible~\cite{Tsukada2017SCRIT}.
Figure~\ref{fig:Sn120_rho} displays the charge and point-neutron density distributions for $^{120}$Sn, a candidate system for the extraction of the symmetry energy slope $L$~\cite{Auma17peelsym}. We evolved the point-proton and point-neutron density operators $\rho_{p/n}(r)$ for each point on a radial mesh from 0 to 10~fm in steps of 0.1~fm. The charge density distribution is then obtained by folding the point-proton and point-neutron distributions with Gaussians following the procedure introduced in~\cite{Chandra1976} and summarized in~\cite{Frosini2021mrII}. The darker theoretical uncertainty bands in Fig.~\ref{fig:Sn120_rho} are obtained by varying the harmonic-oscillator frequency around the optimal value, while the lighter bands show the sensitivity of the results to changing $e_\mathrm{max}$ from 12 to 14. Both bands are an estimate of the remaining model-space uncertainties. For comparison, we also show results based on the NNLOsat interaction~\cite{Ekst15sat}, which proved to reproduce well the charge density distributions in the Sn region~\cite{Arthuis2020a}. For these calculations, we were able to make use of the novel matrix element storage scheme~\cite{Miyagi2021} to yield vastly reduced uncertainties compared to Ref.~\cite{Arthuis2020a}. Results for the charge distribution show a good reproduction of two-point Fermi and three-point Gaussian parametrizations extracted from experiment~\cite{DeVries1987ADNDT}, where differences could only be resolved through a more detailed experimental description. Interestingly, the behaviors obtained for the novel 1.8/2.0~(EM7.5) and 1.8/2.0~(sim7.5) interactions are very similar. Results for the point-neutron density distribution exhibit larger uncertainties, especially in the core of the nucleus. Here the differences with the experimental extraction from Terashima \emph{et al.}~\cite{Terashima2008} are larger, with our \emph{ab initio} results yielding a softer shoulder toward the neutron skin.

\begin{figure}[t!]
  \includegraphics[width=\linewidth]{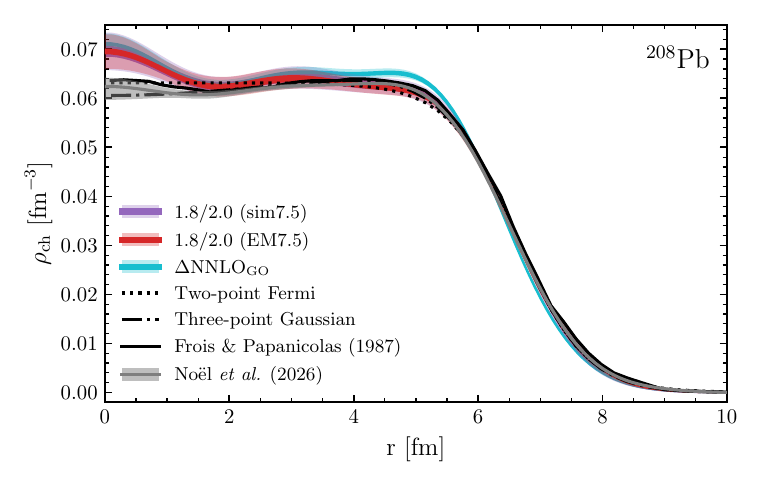}%
  \caption{\label{fig:Pb208_rho_ch} Charge density distribution for $^{208}$Pb for the two novel interactions 1.8/2.0 (EM7.5) and 1.8/2.0 (sim7.5) as well as the $\Delta$NNLO$_\text{GO}$ interaction, compared to data compiled in Refs.~\cite{DeJager1974ADNDT,Frois1987} or recently reanalyzed~\cite{Noel2024,Noel2026}. The theoretical uncertainties are as in Fig.~\ref{fig:Sn120_rho}.}
\end{figure}

Figure~\ref{fig:Pb208_rho_ch} displays the charge density distribution for $^{208}$Pb compared to two-point Fermi and three-point Gaussian parametrizations~\cite{DeJager1974ADNDT}, as well as the compiled data from Frois and Papanicolas~\cite{Frois1987} and analysis with uncertainty quantification by No\"el et al.~\cite{Noel2024,Noel2026}. The model-space convergence uncertainties remain moderate, and overall the reproduction of experiment is very good. Our novel 1.8/2.0~(EM7.5) and 1.8/2.0~(sim7.5) interactions display very similar results, while $\Delta$NNLO$_\mathrm{GO}$ yields a slightly too compressed density with a stronger shoulder at larger distances. Note that convergence with model space cannot be obtained for the NNLOsat interaction for such heavy nuclei. Overall, these results demonstrate the ability of chiral low-resolution interactions to reproduce density distributions over the whole nuclear chart and can provide valuable information for experiments.

\section{Isospin dependence of neutron skins}

\begin{figure}[t!]
  \includegraphics[width=\linewidth]{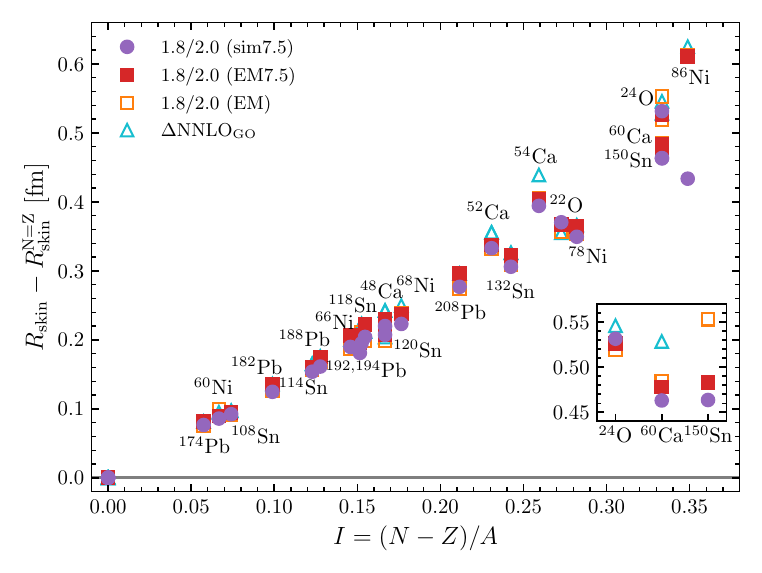}%
  \caption{\label{fig:skins_Coulomb_corrected} Neutron skin with respect to the corresponding $N=Z$ isotope for selected doubly closed-shell nuclei as a function of the isospin asymmetry for the two novel interactions 1.8/2.0 (EM7.5) and 1.8/2.0 (sim7.5) as well as the 1.8/2.0 (EM) and $\Delta$NNLO$_\text{GO}$ interactions.}
\end{figure}

Neutron skins provide a direct connection between nuclei and the equation of state of neutron-rich matter through the symmetry energy slope $L$~\cite{Hu22Pb,Adhikari2021prex2,Essick2021,Reinhard2022,Mammei2023,Birk17dipole,Ross13dipole,Hash15dipole,Tami11dipole,Adhikari2022crex,Auma17peelsym,RocaMaza2015,Fattoyev2018}. Macroscopic models have linked the neutron skin thickness to the isospin asymmetry $I = (N-Z)/A$~\cite{Myers1969,Latt12esymm}, with the correlation being approximately linear. Such a correlation was recently confirmed with \emph{ab initio} calculations for medium-mass nuclei~\cite{Novario2023}. We effectively subtract the Coulomb contribution by plotting $R_\text{skin}$ relative to the (negative) $R_\text{skin}^\text{N=Z}$ of the $N=Z$ isotope in the chain. This is shown in Fig.~\ref{fig:skins_Coulomb_corrected} for selected closed-shell isotopes of O, Ca, Ni, Sn, and Pb. We find that up to $I \approx 0.2$ the growth in neutron skin thickness is linear with respect to $I$ and all interactions yield very similar results, as already apparent for the nuclei studied in Fig.~\ref{fig:Main_selected}. However, for $I \gtrsim 0.25$ nuclei show more scatter and thus a stronger sensitivity on the employed interaction\footnote{In the case of $^{86}$Ni and the 1.8/2.0~(sim7.5) interaction, the significantly smaller neutron skin is associated to an underbinding of the system with respect to other interactions, indicating a possible interaction deficiency for this specific system.}. The inset shows the results for the different nuclei at $I = 1/3$ in more detail.
The interaction $\Delta\mathrm{NNLO}_\mathrm{GO}$ tends to systematically predict larger skins for very neutron-rich nuclei, making exotic neutron-rich nuclei very interesting for the neutron skin but also very challenging.
Moreover, interactions that successfully reproduce radii and neutron skins can open the door to improved constraints for the determination of the neutron-star equation of state~\cite{Drischler2020,Essick2021,Newton2021,Keller2022}.

\section{Conclusions and outlook}

In this paper, we developed novel chiral low-resolution interactions based on SRG-evolved high-precision NN potentials, combined with 3N forces at N$^2$LO fitted to the ground-state energies of $^3$H and $^{16}$O as well as the charge radius of $^{16}$O. The new interactions accurately reproduce experimental ground-state energies, charge radii, and neutron skins for closed-shell nuclei from $^{16}$O up to $^{208}$Pb, and hence resolve shortcomings of the established low-resolution interaction 1.8/2.0 (EM). The new interactions also predict density distributions for $^{120}$Sn and $^{208}$Pb in remarkable agreement with experiment. Finally, we investigated neutron skins for neutron-rich nuclei, a key quantity that links nuclei to infinite nuclear matter and astrophysics. We found that neutron skins are narrowly predicted over all nuclei with interesting sensitivities for the most extreme, experimentally unexplored cases. This motivates further investigations of neutron-rich nuclei with rare isotope beam facilities in order to obtain constraints for nuclear forces and to better understand the connection to the equation of state. Our novel low-resolution interactions present a powerful tool for comprehensive \emph{ab initio} studies of the nuclear chart, including open-shell nuclei as well as other observables like excited states or response functions to external probes. First results for excitation energies are given in \ref{app:twoplus}.

\section*{Acknowledgements}

We thank T.~Aumann, M.~Heinz, T.~Miyagi, and A.~Tichai for helpful discussions.
Three-body matrix elements of the interactions were generated using the NuHamil code~\cite{Miyagi2023}, and calculations using the IMSRG code from S.~R.~Stroberg~\cite{Stro17imsrggit}.
This work was supported in part by the European Research Council (ERC) under the European Union's Horizon 2020 research and innovation programme (Grant Agreement No.~101020842), the Deutsche  Forschungsgemeinschaft (DFG, German Research Foundation) -- Project-ID 279384907 -- SFB 1245, and the Helmholtz Forschungsakademie Hessen für FAIR (HFHF).
The authors gratefully acknowledge the Gauss Centre for Supercomputing e.V.~(www.gauss-centre.eu) for funding this project by providing computing time through the John von Neumann Institute for Computing (NIC) on the GCS Supercomputer JUWELS at J\"ulich Supercomputing Centre (JSC).

\appendix

\section{Incorporation of many-body observables}
\label{app:manybody}

\begin{table}[b!]
\caption{\label{tab:LECs} 3N LECs for the 1.8/2.0 (EM) interaction~\cite{Hebe11fits} (left column) and the two novel low-resolution NN+3N interactions (middle and right columns). The interaction 1.8/2.0 (EM7.5) is based on the same N$^3$LO EM 500 NN potential~\cite{Ente03EMN3LO} while 1.8/2.0 (sim7.5) is based on the NNLOsim NN potential~\cite{Carl15sim} for $T^\mathrm{max}_\mathrm{Lab} = 290$\,MeV and $\Lambda = 550$\,MeV.}

\begin{tabularx}{\columnwidth}{@{}cccc@{}}
    \hline\hline
    LEC & 1.8/2.0 (EM) & 1.8/2.0 (EM7.5) & 1.8/2.0 (sim7.5) \vphantom{\Large H} \\\hline
    $c_1$ & $-0.81$ & $-0.81$ & 0.27  \\
    $c_3$ & $-3.2$ & $-3.2$ & $-3.56$  \\
    $c_4$ & 5.4 & 5.4 & 3.644  \\
    $c_D$ & 1.264 & 7.5 & 7.5 \\
    $c_E$ & $-0.12$ & 0.942  & 1.081  \\
    \hline\hline
\end{tabularx}
\end{table}

\begin{figure}[t!]
  \includegraphics[width=\linewidth]{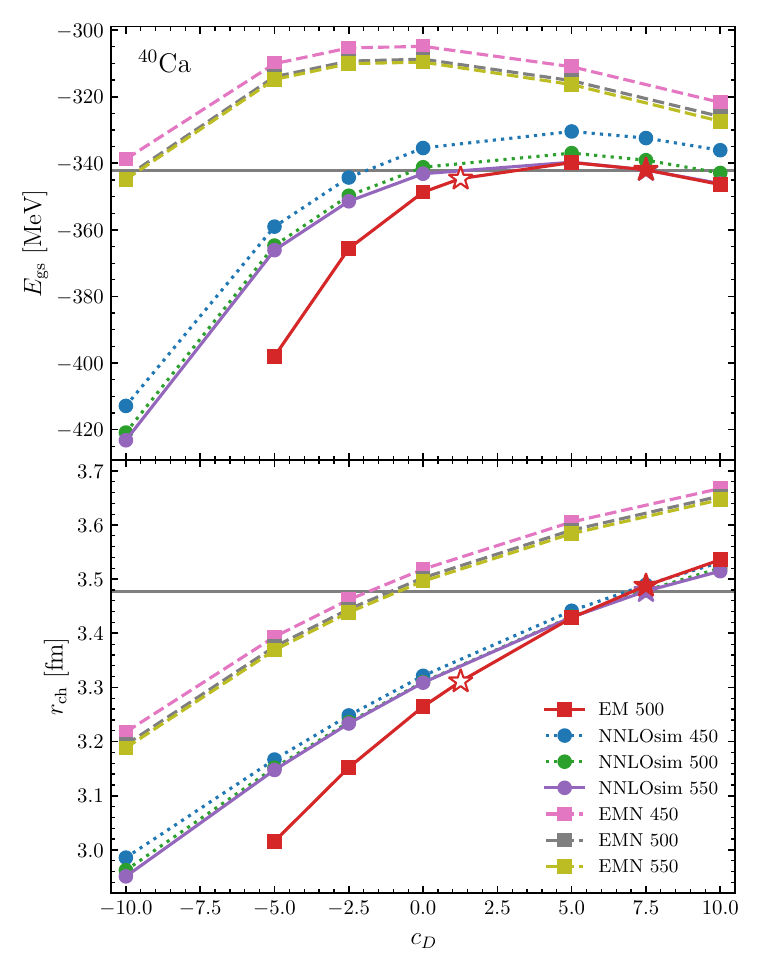}%
  \caption{\label{fig:Ca40_Egs} Ground-state energy (top panel) and charge radius (bottom panel) of $^{40}$Ca as a function of $c_D$ for the low-resolution NN+3N interactions shown in Fig.~\ref{fig:3Hfits}. Horizontal grey lines correspond to the experimental values~\cite{Wang2021AME,Ange13rch}. The solid stars indicate the results corresponding to the two novel interactions 1.8/2.0 (EM7.5) and 1.8/2.0 (sim7.5), while the open star corresponds to the 1.8/2.0 (EM) interaction~\cite{Hebe11fits}.}
\end{figure}

Starting from the correlations shown in Fig.~\ref{fig:3Hfits}, we fix the values of the LECs based on the ground-state energy and charge radius of $^{16}$O. Figure~\ref{fig:O16_Egs} shows the predicted results as a function of $c_D$ for the different 1.8/2.0 interactions. We find that for the NN potential EM 500 and for the NNLOsim family, evolved to the resolution scale $\lambda=1.8 \, \text{fm}^{-1}$, it is possible to obtain a simultaneous fit of the $^3$H and $^{16}$O ground-state energies as well as the $^{16}$O charge radius using chiral N$^2$LO 3N interactions with $\Lambda_{\text{3N}} = 2.0 \, \text{fm}^{-1}$. The corresponding LECs of the 3N interactions are summarized in Tab.~\ref{tab:LECs}. Figure~\ref{fig:Ca40_Egs} shows that the corresponding results also hold for $^{40}$Ca. The fact that the low-resolution EMN interactions do not allow for a simultaneous fit of all three observables can be traced back to the $c_i$ values. In particular, we observed that adjustments of $c_4$ mainly affect the curvature of the parabola observed for both the $^{16}$O and $^{40}$Ca ground-state energy as a function of $c_D$, while $c_1$ and $c_3$ mostly lead to constant shifts of the energy and charge radius results. In addition, we also studied the sensitivity of the results to the shape of the 3N interaction regulator function $f(p,q) = \exp[ - ((p^2 + \frac{3}{4} q^2)/\Lambda_{\text{3N}}^2)^{n_\mathrm{exp}}]$ (with the Jacobi momenta $p$ and $q$~\cite{Hebe203NF}), which is effectively controlled by $n_\mathrm{exp}$. Different values have been employed for this parameter in recent works: While in Ref.~\cite{Hebe11fits} the value $n_{\mathrm{exp}}$ = 4 was used, in Refs.~\cite{Carl15sim,Huth19chiralfam} $n_{\mathrm{exp}}$ = 3 was used. In this work, we use $n_{\mathrm{exp}} = 4$, except for the NNLOsim family, for which we kept the original choice  $n_{\mathrm{exp}} = 3$. However, investigating both values for NNLOsim, we found that these different choices have only a very small impact on the results and hence cannot explain the difference between the interactions.

\section{Properties of few-body systems}
\label{app:fewbody}

\begin{table*}[t!]
\caption{\label{tab:fewbody} $^3$H and $^4$He ground-state energies and radii computed for the 1.8/2.0 interactions by solving the Faddeev equations ($^3$H) and the Jacobi no-core-shell model ($^4$He), in comparison with experiment~\cite{Wang2021AME,Ange13rch}.}
\begin{center}
\begin{tabular}{ccccc}
    \hline \hline
    {} & 1.8/2.0 (EM) & 1.8/2.0 (EM7.5) & 1.8/2.0 (sim7.5) & Exp. \\\hline
    $E_\mathrm{gs}(^3\mathrm{H})$ [MeV] & $ -8.48 $ & $ -8.48 $ & $ -8.48 $ & $-8.482$ \\
    $r_\mathrm{p}(^3\mathrm{H})$ [fm] & $ 1.62 $ & $ 1.63 $ & $ 1.63 $ & $1.608$ \\
    $E_\mathrm{gs}(^4\mathrm{He})$ [MeV] & $-29.11$ & $-27.99$ & $-27.65$ & $-28.296$ \\
    $r_\mathrm{p}(^4\mathrm{He})$ [fm] & $1.45$ & $1.52$ & $1.53$ & $1.478$ \\
    \hline \hline
\end{tabular}
\end{center}
\end{table*}

\begin{table*}[t!]
\caption{\label{tab:N02B} $^4$He ground-state energy and point-proton radius computed from the 1.8/2.0 interactions at the IMSRG(2) level compared to the Jacobi no-core shell model.}
\begin{center}
\begin{tabular}{ccccccc}
    \hline \hline
    {} & \multicolumn{2}{c}{1.8/2.0 (EM)} & \multicolumn{2}{c}{1.8/2.0 (EM7.5)} & \multicolumn{2}{c}{1.8/2.0 (sim7.5)} \\
      {}  & IMSRG(2) & NCSM & IMSRG(2) & NCSM & IMSRG(2) & NCSM \\\hline
    $E_\mathrm{gs}(^4\mathrm{He})$ [MeV] & $-29.24$ &  $-29.11$ & $-28.71$ & $-27.99$ & $-28.03$ & $-27.65$ \\
    $r_\mathrm{p}(^4\mathrm{He})$ [fm] & $1.43$ & $1.45$ & $1.56$ & $1.52$ & $1.55$ & $1.53$  \\
    \hline \hline
\end{tabular}
\end{center}
\end{table*}

In Table~\ref{tab:fewbody} we summarize results for ground-state properties of $^3$H and $^4$He obtained from solving Faddeev equations ($^3$H) and using the Jacobi no-core shell model ($^4$He) for all considered 1.8/2.0 interactions. The ground-state energy of $^3$H is reproduced exactly in all cases by construction, and all interactions yield a similar, slightly too large value of the point-proton radius for this nucleus. The experimental values of the point-proton radius are obtained from the charge radius values of Ref.~\cite{Ange13rch} as follows, $r_\mathrm{p}^2 = r_\mathrm{ch}^2 - R_\mathrm{p}^2 - (N/Z) R_\mathrm{n}^2 - r_\mathrm{DF}^2,
$ where $r_\mathrm{DF}^2 = 0.332$ is the Darwin-Foldy correction, $R_\mathrm{p} = 0.8409\ \mathrm{fm}$ is the proton radius and $R_\mathrm{n}^2 = -0.1161\ \mathrm{fm}^2$~\cite{Zyla2020} the neutron radius squared. These values are consistent with the ones used for the prediction of charge radii in the present work, where the spin-orbit correction is also taken into account. For $^4$He, the 1.8/2.0~(EM) interaction reproduces the point-proton radius almost exactly by construction, although based on older values of the proton and neutron radii~\cite{Hebe11fits}, and predicts a slightly too bound nucleus. For both the 1.8/2.0~(EM7.5) and 1.8/2.0~(sim7.5) interactions, we obtain a small underbinding and too large radii for $^4$He. This shows that an excellent reproduction of $^3$H is not sufficient to obtain a good reproduction of the properties of $^4$He.

Table~\ref{tab:N02B} displays predictions for $^4$He based on the 1.8/2.0 interactions using the IMSRG(2) and the Jacobi no-core shell model. Overall, we observe a reasonable agreement between the IMSRG(2) results and the exact results of the no-core shell model. This is similar for other interactions and is not only due to the normal-ordered two-body approximation but also to the IMSRG(2) truncation discarding induced three- and four-body forces during the IMSRG flow. In particular, although the deviation of about 700\,keV for the 1.8/2.0~(EM7.5) interaction might seem larger at first, the normal-ordered two-body approximation has been shown not to perform as well for $^4$He~\cite{Roth12NCSMCC3N,Simo17SatFinNuc} compared to medium-mass nuclei where it typically yields errors below 1\%~\cite{Bind13expl3NCCSD}.

\section{Excitation energies}
\label{app:twoplus}

\begin{figure}[t!]
    \includegraphics[width=\linewidth,clip=]{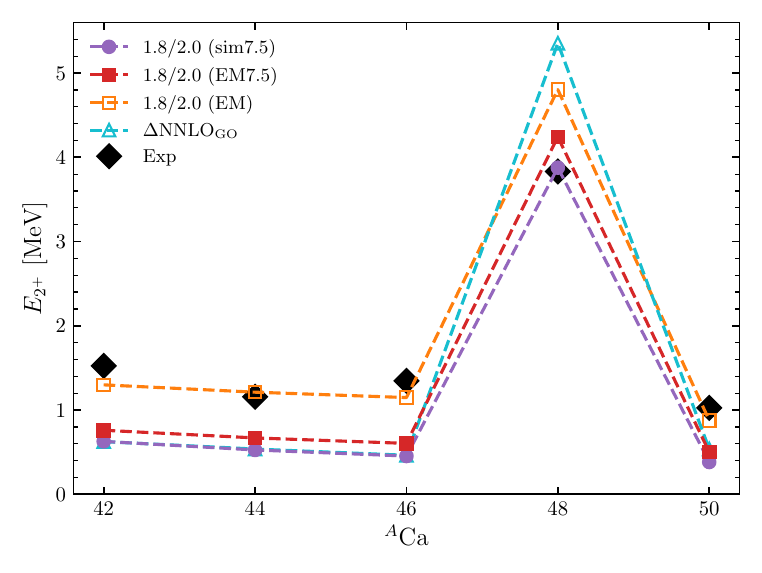}
    \caption{\label{fig:Ex_Ca} Excitation energy of the first $2^+$ state of even Ca isotopes for the two novel interactions 1.8/2.0 (EM7.5) and 1.8/2.0 (sim7.5) as well as the 1.8/2.0 (EM) and $\Delta$NNLO$_\text{GO}$ interactions compared to experiment~\cite{Chen2016,Chen2023a,Wu2000,Chen2022,Chen2019}.}
\end{figure}

As a first test of the performance of the new interactions for spectra, we present results for the excitation energy of the first $2^+$ state of the even $^{42-50}$Ca isotopes in Fig.~\ref{fig:Ex_Ca}. Calculations are performed using the valence-space IMSRG~\cite{Stro17ENO} using the KSHELL code~\cite{kshell}, with a $0\hbar\omega$-valence space, i.e., with active neutrons in the $pf$-shell on top of a $^{40}$Ca core.

We observe that the two new 1.8/2.0~(EM7.5) and 1.8/2.0~(sim7.5) interactions underpredict the excitation energies of open-shell isotopes away from $^{48}$Ca, similarly to the $\Delta\mathrm{NNLO}_\mathrm{GO}$ interactions. The subshell closure for $^{48}$Ca is qualitatively reproduced by all interactions. The agreement of the new interactions with the experimental value for $^{48}$Ca is presumably accidental, as missing correlations from higher many-body contributions are expected to reduce the excitation energy~\cite{Hage16Ni78}, but this needs to be explored for the new interactions.

\bibliographystyle{apsrev4-1}
\bibliography{strongint}

\end{document}